\documentclass[12pt]{iopart}
\usepackage[sort,compress,comma,numbers]{natbib}
\setcitestyle{square}
\setcitestyle{citesep={,}}
\usepackage{graphicx}
\usepackage{setstack}
\usepackage{subfig}
\usepackage{soul,xcolor}

\usepackage{iopams}

\expandafter\let\csname equation*\endcsname\relax
\expandafter\let\csname endequation*\endcsname\relax

\expandafter\let\csname subarray\endcsname\relax
\expandafter\let\csname endsubarray\endcsname\relax

\expandafter\let\csname ddddot\endcsname\relax
\expandafter\let\csname ddddot\endcsname\relax

\expandafter\let\csname dddot\endcsname\relax
\expandafter\let\csname dddot\endcsname\relax

\expandafter\let\csname uproot\endcsname\relax
\expandafter\let\csname uproot\endcsname\relax

\expandafter\let\csname leftroot\endcsname\relax
\expandafter\let\csname leftroot\endcsname\relax

\usepackage{amsmath}

\begin{document}

\title[Semiclassical dynamics of a superconducting circuit]{Semiclassical dynamics of a superconducting circuit: chaotic dynamics and fractal attractors}

\author{Davide Stirpe$^{1,2}$, Juuso Manninen$^1$, Francesco Massel$^1$}
\address{$^1$ Department of Science and Industry Systems, University of South-Eastern Norway, PO Box 235, Kongsberg, Norway}
\address{$^2$ Politecnico di Torino, Corso Duca degli Abruzzi 24, I–10129 Torino, Italy}
\ead{francesco.massel@usn.no}


\begin{abstract}
We study here the semiclassical dynamics of a superconducting circuit constituted by two Josephson junctions in series, in the presence of a voltage bias. We derive the equations of motion for the circuit through a Hamiltonian description of the problem, considering the voltage sources as semi-holonomic constraints. We find that the dynamics of the system corresponds to that of a planar rotor with an oscillating pivot. We show that the system exhibits a rich dynamical behaviour with chaotic properties and we present a topological classification of the cyclic solutions, providing insight into the fractal nature of the dynamical attractors.
\end{abstract}

\section{\label{sec:intro}Introduction}

Superconducting circuits are arguably one of the most successful quantum computing platforms~\cite{Wendin.2017}. At the heart of these circuits is the Josephson junction (JJ), which, in a lumped-elements description, acts as a nonlinear circuit element~\cite{Josephson.1962,girvin2014circuit}. In a quantum setting, this nonlinearity allows isolating a two-level system within the full spectrum, and, consequently, constructing a wide collection of superconducting qubits~\cite{Mooij.1999,Martinis.2002,Koch.2007}, depending on the specific topology of the circuit. Beyond the field of quantum computation, the physics of JJs has also been discussed and applied in other contexts, such as the study of quantum phase transitions in condensed matter physics~\cite{Martinoli.2000,Cataliotti.2001}, sensing and metrology applications~\cite{Fagaly.2006}, where striking consequences of the tunnelling between superconductors can already be identified at the semiclassical level. One example is given by the Aharonov-Bohm effect~\cite{Aharonov.1959} in SQUIDS. These devices, consisting of a superconducting loop interrupted by two dielectric layers and modelled as two JJs in parallel, exploit the relation between the superconducting phase and the magnetic flux to act as extremely sensitive magnetometers.

In this article, we focus on this semiclassical aspect: in a circuit somewhat analogous to the SQUID, we describe the dynamics of the superconducting phase of a device constituted, unlike the SQUID, by two JJs in series in the presence of an external voltage bias $V_\mathrm{g}$ (see figure~\ref{series}a). This setup defines a superconducting island coupled to the rest of the circuit through two (superconducting) tunnel junctions. As we will discuss in greater detail below, we note that, even if this circuit has the same topology of a superconducting single-electron transistor (SSET) at zero gate voltage~\cite{Pekola.2013}, our analysis is valid in a regime which is complementary to the one required to observe charge quantization in SSETs.

The dynamics of the circuit is discussed in terms of Hamiltonian dynamics, allowing us to recognize how the equation of motion for the phase on the island between the JJs (see figure~\ref{series}) can be mapped onto the dynamics of an equivalent mechanical system: a parametrically driven planar rotor (a planar rotor whose pivot is periodically driven). While the correspondence between the dynamics of a \emph{single} JJ and a driven physical pendulum has been extensively studied in the literature~\cite{zhang2011quantitative,gwinn1986fractal,sobolewski1988chaos,kautz1996noise,iansiti1985noise,macdonald1983study}, the relation between the series of two JJs and a mechanical system is, to our knowledge, novel. 

The interest in the parametrically driven planar rotor lies in the fact that it can be considered as a zero-gravity parametrically driven pendulum (PDP). The latter is a system with a rich dynamical behaviour, exhibiting a wide collection of attractors and fixed points, such as the non-inverted ($\theta = 0$) and inverted ($\theta = \pi$) positions. In particular, the inverted position can be stabilized for certain values of the system parameters~\cite{Stephenson.1908,Kapitza.1951,clifford1998inverted,kim1998bifurcations} --a configuration often referred to as the Kapitza pendulum~\cite{Butikov.2011}. The general stability properties of the PDP have been studied in several works~\cite{acheson1995multiple,blackburn1995stochastic,clifford1998inverted,kim1998bifurcations,bartuccelli2001dynamics,bartuccelli2002stability,xu2005rotating,butikov2005complicated,Carbo.2010,Butikov.2011}, considering gravity, driving amplitude and, in some instances, friction as free parameters. While the analytical investigation has mainly focused on the stability of the inverted position in terms of solutions of the Mathieu equation~\cite{bartuccelli2002stability}, numerical work has addressed various aspects of the PDP dynamics. These include the appearance of nontrivial limit-cycles (multiple nodding solutions~\cite{acheson1995multiple,butikov2005complicated}), the emergence of fractal basins of attraction~\cite{bartuccelli2001dynamics}, and the investigation of "rotating solutions" (i.e. unbounded rotations)~\cite{xu2005rotating}.

In our work, we first discuss the Hamiltonian formalism used to derive the equations of motion of the electrical degrees of freedom of the circuit. We start from the Lagrangian associated with the lumped-elements description of the device, in which the external voltage source is considered as a semi-holonomic constraint on the local voltages. This allows us to identify the dynamics of the circuit of figure~\ref{series}a with that of the parametrically driven planar rotor in figure~\ref{series}b. Having in mind our circuit implementation, we then discuss several new aspects of the PDP. First, we distinguish between unstable, $0$-stable, $\pi$-stable and limit-cycle asymptotic attractors for the superconducting phase of the circuit $\varphi_\Delta (t)$. By focusing on the natural conditions for the JJ setting ($\dot{\varphi}_\Delta (0) = 0$), we provide a description of the fractal nature of the stability diagram obtained by varying the dimensionless driving amplitude and the initial superconducting phase: we compute the Hausdorff dimension of the attractors. Moreover, we classify the limit-cycle solutions (n-cycles) in terms of the number of intersections with the zero-velocity axis in the phase space, providing then a topological classification of these trajectories. We find that the system displays a wide collection of n-cycles characterized by a chaotic distribution in the stability diagram, identifying cyclic solutions having up to $n = 18$. Furthermore, we analyze the case $\dot{\varphi}_\Delta (0) \ne 0$ for two fixed values of the dimensionless driving amplitude, obtaining basins of attraction with fractal geometry, whose borders have non-integer Hausdorff dimension. 

\section{\label{sec:2}Semiclassical model and equations of motion}
\begin{figure}[htb!]
   \centering
   \captionsetup{justification=raggedright}
    \includegraphics[width=15cm]{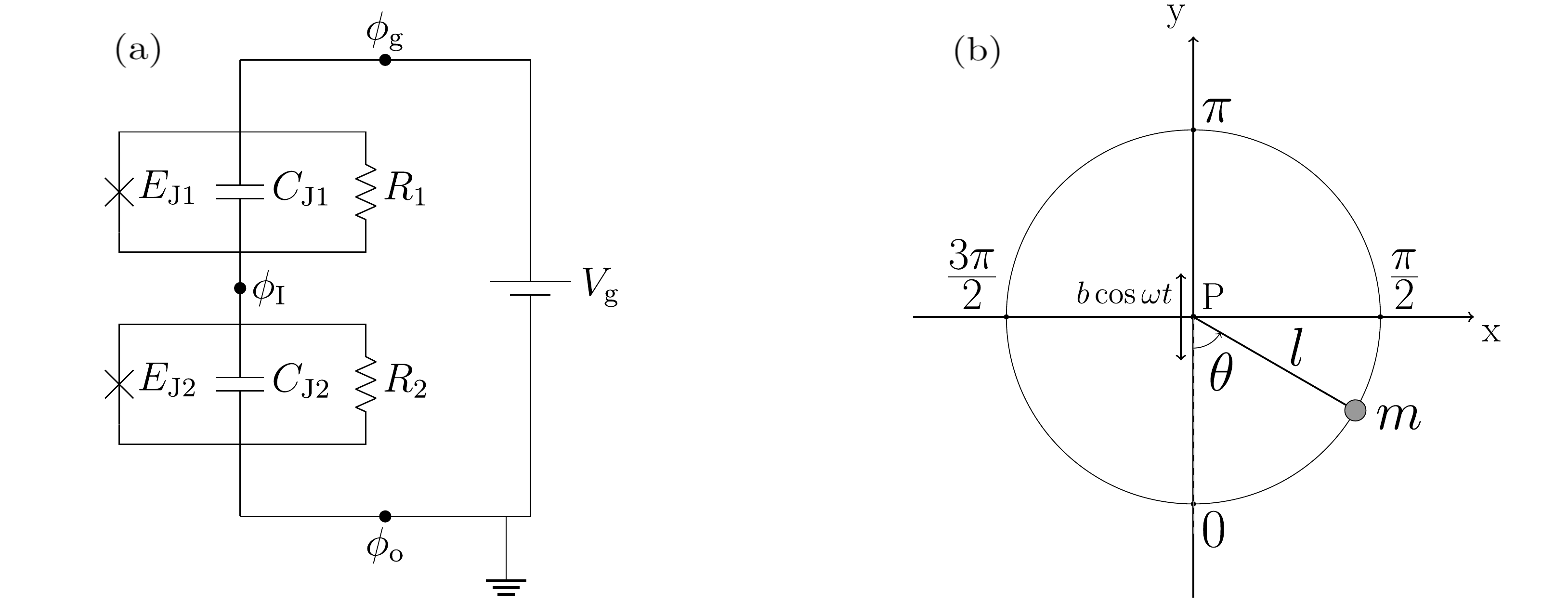}
    \caption{(a) Equivalent circuit of the superconducting device. The Josephson junction consists of capacitances $C_{\mathrm{J}1}$, $C_{\mathrm{J}2}$ and  non-linear inductances $E_{\mathrm{J}1}$, $E_{\mathrm{J}2}$ in parallel. We introduce shunting resistors $R_1$,$R_2$ to take into account dissipation. The system is voltage biased ($V_\mathrm{g}$). (b) Schematic representation of a parametrically driven pendulum, consisting of a rod of length $l$ with a pivot P oscillating in the vertical direction and a mass $m$ attached to the free end.}
    \label{series}
\end{figure}

The system considered is a superconducting circuit constituted by two JJs in series, in the presence of a voltage bias $V_\mathrm{g}$ (figure~\ref{series}a). In a lumped-elements circuit description, the two Josephson junctions are characterized by capacitances $C_{\mathrm{J}1}$ and $C_{\mathrm{J}2}$ and Josephson energies $E_{\mathrm{J}1}$ and $E_{\mathrm{J}2}$, respectively. While we initially focus on a purely reactive circuit model, we later introduce dissipation by considering a shunt resistance for each of the two JJs, using the RCSJ model~\cite{Stewart.1968}. This resistive shunt can, for instance, describe a quasiparticle contribution to the tunnelling current.

The circuit we consider here has the same topology of a SSET~\cite{Pekola.2013}. The key distinction between our setting and a SSET is that, for the latter, the focus is on the so-called Coulomb blockade regime~\cite{Pekola.2013}, corresponding to charge quantization on the island. Such a regime is reached for a characteristic impedance larger than the resistance quantum  ($Z \gg R_{K}=h/e^2$). Here, we focus on the opposite situation ($Z \ll R_K$), in analogy to the strategy employed in the design of phase qubits~\cite{Wendin.2017, steffen2006state, Martinis.2002}. This can be achieved by allowing the tunnelling junctions to be large, even macroscopic, increasing their capacitance and thus lowering the charging energy $E_\mathrm{C} \propto 1/C$ or, alternatively, adding a large shunting capacitance: incidentally, we note that these conditions suggest that the experimental realization of this device should not pose any particular challenge. A consequence of this choice of parameters is that the number of Cooper pairs on the island is no longer a good quantum number ---contrary to the SET case--- and, instead, the phase dynamics becomes the dominating effect, allowing us to treat the superconducting phase across the junctions semiclassically.

The dynamics of the electrical degrees of freedom is obtained through a Hamiltonian mechanics description of our circuit, starting from a lumped-elements model. In this approach, node fluxes and charges (time integrals of voltages and currents, respectively) are the conjugate variables that play the same role as position and momentum in a conventional classical mechanics problem. Our approach deviates from the standard method of describing the lumped-element model of a superconducting circuit~\cite{girvin2014circuit, vool2017introduction} in that we directly include the voltage constraint imposed by the voltage source through the undetermined Lagrange multiplier method. To our knowledge, this approach has not been considered for the analysis of (superconducting) circuits, where, typically, the voltage source is replaced by capacitors which, in the limit of infinite capacitance, induce the correct node voltage provided by the voltage sources~\cite{girvin2014circuit, vool2017introduction}. We note that our analysis allows for a straightforward quantization procedure which, however, is not the focus of our work.

The general approach consists in identifying the nodes in the circuit under consideration ($i \in\left\{g,I,o\right\}$, in our case) and introducing flux variables $\phi_i$ at each of them. The flux $\phi_i$  is defined as $\phi_i(t) = \int_{-\infty}^t dt' V_i(t')$ where $V_i$ is the voltage at each node, implying that $V_i(t) = \dot{\phi}_i (t)$. The voltage source imposes the constraint $\dot{\phi}_\mathrm{g} - \dot{\phi}_\mathrm{o} - V_\mathrm{g} = 0$ on the circuit. Without loss of generality, we choose here the node $o$ to be grounded, i.e. $V_\mathrm{o} = 0$ and hence $\dot{\phi}_\mathrm{o} = 0$.
The Lagrangian of the superconducting circuit can therefore be written as
\begin{equation}
\mathcal{L} =\mathcal{L}(\vec{\phi}, \vec{\dot{\phi}}) 
    = \frac{C_{\mathrm{J}1}}{2} \left( \dot{\phi}_\mathrm{g} - \dot{\phi}_\mathrm{I} \right)^2 
    + \frac{C_{\mathrm{J}2}}{2} \dot{\phi}_\mathrm{I}^2
    - \mathcal{U}(\vec{\phi}) 
    + Q \left( \dot{\phi}_\mathrm{g} - V_\mathrm{g} \right) ,
\label{lagrangian}
\end{equation}
where $\hspace{0.1 cm}\mathcal{U}(\vec{\phi}) = -E_{\mathrm{J}1}\cos\left[2\pi (\phi_\mathrm{g} - \phi_\mathrm{I})/\Phi_0 \right] - E_{\mathrm{J}2}\cos\left[2\pi \phi_\mathrm{I}/\Phi_0 \right]$ with $\Phi_0 = h/(2e)$ is the inductive energy associated to the two JJs. The rightmost term in the Lagrangian consists of the undetermined Lagrange (charge) multiplier $Q$ and the semi-holonomic (function of the flux derivatives) constraint $\dot{\phi}_\mathrm{g} - \dot{\phi}_\mathrm{o} - V_\mathrm{g} = 0$ on the voltages. Even if the constraint is non-holonomic, its linearity in the derivatives $\vec{\dot{\phi}}$ ensures that the Lagrange multiplier method can be applied (see~\ref{app1}).

The node charges $Q_i = \partial \mathcal{L}/\partial \dot{\phi}_i$, representing the momenta conjugated to the fluxes $\phi_i$, allow us to write the Hamiltonian $\mathcal{H} = \sum_{i}Q_i \dot{\phi}_i - \mathcal{L}$ as
\begin{equation}
\begin{split}
    \mathcal{H} =& 4 \frac{e^2}{2 C_{\mathrm{J}1}} \left(n_\mathrm{g} - N \right)^2
        + 4 \frac{e^2}{2 C_{\mathrm{J}2}} \left(n_\mathrm{I} + n_\mathrm{g} - N \right)^2 
        + 2 e N V_\mathrm{g} \\
        &- E_{\mathrm{J}1}\cos\left(\varphi_\mathrm{g} - \varphi_\mathrm{I}\right)
        -E_{\mathrm{J}2}\cos\left(\varphi_\mathrm{I}\right),
\end{split}
\label{hamiltonian}
\end{equation}
where the charges are now written in terms of the number of Cooper pairs $n_i = Q_i/(2e)$ and the Lagrange multiplier as $N = Q/(2e)$; furthermore, the phases $\varphi_\mathrm{i} = 2\pi\phi_\mathrm{i} /\Phi_0$ are introduced (flux-phase relation).
From equation~\eqref{hamiltonian}, we can write Hamilton's equations for the node variables as
\begin{subequations}
\begin{align}
    \dot{n}_\mathrm{I} &= - \frac{1}{\hbar} \frac{\partial \mathcal{H}}{\partial \varphi_\mathrm{I}}
        = -\frac{1}{\hbar} \left[ E_{\mathrm{J}1} \sin(\varphi_\mathrm{I} - \varphi_\mathrm{g})
            + E_{\mathrm{J}2} \sin(\varphi_\mathrm{I}) \right], \label{nI} \\
    \dot{n}_\mathrm{g} &= - \frac{1}{\hbar} \frac{\partial \mathcal{H}}{\partial \varphi_\mathrm{g}}
        = \frac{1}{\hbar} E_{\mathrm{J}1} \sin(\varphi_\mathrm{I} - \varphi_\mathrm{g}), \label{nG} \\
    \dot{\varphi}_\mathrm{I} &= \frac{1}{\hbar} \frac{\partial \mathcal{H}}{\partial n_\mathrm{I}}
        =  \frac{8}{\hbar} \frac{e^2}{2 C_{\mathrm{J}2}}(n_\mathrm{I} + n_\mathrm{g} - N), \label{phiI} \\
    \dot{\varphi}_\mathrm{g} &= \frac{1}{\hbar} \frac{\partial \mathcal{H}}{\partial n_\mathrm{g}}
        =  \frac{8}{\hbar} \left[ \frac{e^2}{2 C_{\mathrm{J}2}} n_\mathrm{I} 
            + \left( \frac{e^2}{2 C_{\mathrm{J}1}} + \frac{e^2}{2 C_{\mathrm{J}2}} \right) 
            \left( n_\mathrm{g} - N \right) \right]. \label{phiG}
\end{align}
\end{subequations}
Using the voltage constraint together with equation~\eqref{phiG}, one can solve for the Lagrange multiplier
\begin{equation}
\label{eq:Lagrangian_multiplier}
    N = \frac{1}{2e}\frac{C_{\mathrm{J}1}C_{\mathrm{J}2}}{\left(C_{\mathrm{J}1} + C_{\mathrm{J}2}\right)} V_{\rm g},
\end{equation}
realizing that it corresponds to the number of Cooper pairs provided by the external voltage source to the two JJ capacitances $C_{\mathrm{J}1}$ and $C_{\mathrm{J}2}$  in series, in absence of tunnelling across the junctions.

In the following, we will assume that $E_{\mathrm{J}1} = E_{\mathrm{J}2} = E_\mathrm{J}$ and define $E_\mathrm{C} = e^2/\left[2(C_{\mathrm{J}1} + C_{\mathrm{J}2})\right]$. Differentiating equation~\eqref{phiI} with respect to time and substituting equations~\eqref{nI} and \eqref{eq:Lagrangian_multiplier} in it, we obtain
\begin{equation}
\label{eq:Kapitza_EOM}
    \ddot{\varphi}_\Delta = - 2 \Omega_\mathrm{J}^2 \cos \left( \frac{\varphi_\mathrm{g}(t)}{2} \right) \sin \left( \varphi_\Delta \right),
\end{equation}
where $\varphi_\Delta = \varphi_\mathrm{I} - \varphi_\mathrm{g}/2$, $\varphi_{\mathrm{g}}(t) =  2\pi V_{\mathrm{g}}/\Phi_0\,t$ (we impose $\varphi_{\mathrm{g}}(0) = 0$) and $\Omega_\mathrm{J} =\sqrt{8E_{C}E_\mathrm{J}}/\hbar$ is the Josephson frequency. Equation \eqref{eq:Kapitza_EOM} can be written in a dimensionless form as
\begin{equation}
\label{EQ:JJs_adimensional}
    \partial_{\rm \tau}^2 \varphi_\Delta
        = - \bar{\epsilon}\cos\left(\tau\right)\sin{\varphi_\Delta},
\end{equation}
where $\bar{\epsilon}=8\Omega_\mathrm{J}^2/\omega_\mathrm{g}^2$, $\tau = \omega_\mathrm{g} t/2$ and $\omega_\mathrm{g} = 2\pi V_\mathrm{g}/\Phi_0$. See~\ref{app1} and~\ref{app2} for further details on the derivation of the system's Hamiltonian and the equations of motion. Introducing dissipation, equation~\eqref{EQ:JJs_adimensional} becomes
\begin{equation}
    \partial_\tau^2\varphi_\mathrm{\Delta} = -\bar{\epsilon}\cos\left(\tau\right)\sin{\varphi_{\Delta}} -\kappa \partial_{\rm \tau }\varphi_{\mathrm{\Delta}},
    \label{kapitza2}
\end{equation}
where the (dimensionless) dissipation term is given by $\kappa = 4/\left[\omega_\mathrm{g}R (C_{\mathrm{J}1}+C_{\mathrm{J}2})\right]$ and $R$ describes the resistance used to model the losses within the RCSJ model: see~\ref{app3} for a detailed derivation. 

In our analysis, we will focus on the dynamics generated by equation~\eqref{kapitza2} for different values of the parameter $\bar{\epsilon}$ and the initial conditions $\left[\varphi_\Delta (0), \dot{\varphi}_\Delta (0)\right]$. In terms of our device, this choice of the dynamical parameters appears natural: the parameter $\bar{\epsilon}$ is controlled by the voltage $V_\mathrm{g}$; the initial condition on the phase $\varphi_\Delta (0)$ corresponds to introducing a time-independent magnetic flux $\Phi_B/\Phi_0=2  \varphi_\Delta(0)$ for $t<0$ through the main loop of the circuit. The initial condition of the phase derivative $\dot{\varphi}_\Delta (0) = 0$ indicates a symmetric voltage drop across the JJs, i.e. $\dot{\varphi}_{\rm I} (0) = V_\mathrm{g} / 2$. An initial condition $\dot{\varphi}_\Delta (0) \ne 0$ corresponds therefore to an initially asymmetric voltage drop across the junction $V_{1,2} = V_\mathrm{g} / 2 \mp \dot{\phi}_\Delta (0)$, where $V_1$ ($V_2$) is the voltage drop across the first (second) JJ.

It is possible to recognize that equation~\eqref{kapitza2} is a specific case of the equation controlling the PDP dynamics. The latter consists of a rigid planar rotor of length $l$ whose pivot is driven harmonically along the vertical direction as
\begin{equation}
  z_{\rm pivot}=b \cos(\omega t).
  \label{eq:pivot}
\end{equation}
Following~\cite{bartuccelli2001dynamics}, the equation of motion for the angle can be written as 
\begin{equation}
  \label{EQ:kapitza_adimensional}
  \partial_\tau^2 \theta =- \left[\alpha - \beta \cos\left(\tau\right)  \right]\sin\left(\theta \right) - \gamma \partial_{\tau}\theta,
\end{equation}
where $\beta=b/l$, $\alpha=g/(l \omega^2)$ and $\tau=\omega t$, and $\gamma$ is a coefficient modelling dissipation. Setting $\beta=\bar{\epsilon}$ and $\alpha=\left[g/(l \omega^2)\right]=0$, we can identify the equation controlling the dynamics of our system with the PDP equations of motion for $g=0$.

Interestingly, in the limit $V_\mathrm{g} = 0$ ($\bar{\epsilon}\rightarrow \infty$), equation \eqref{eq:Kapitza_EOM} reduces to the case of a physical (i.e. not parametrically driven) pendulum in the presence of gravity, whose value is determined by the Josephson frequency and the initial superconducting phase. From the PDP perspective, the $V_\mathrm{g} = 0$ condition in the JJ circuit translates into a constant acceleration of the pivot and therefore, into an apparent force applied to the oscillating mass, caused by the non-inertiality of the reference frame. The $V_\mathrm{g} = 0$ limit is not investigated in this work any further.

\section{\label{sec:4}Numerical results}

In this section, we perform a numerical analysis of equation~\eqref{kapitza2} with two different settings of the parameters. In one case, we vary the dimensionless driving amplitude $\bar{\epsilon}$ and the initial phase $\varphi_\Delta (0)$, fixing $\dot{\varphi}_\Delta (0) = 0$. In the other case, we pick up two values of $\bar{\epsilon}$ and we vary both the initial conditions $\left[\varphi_\Delta(0), \dot{\varphi}_\Delta(t)\ne 0\right]$. In both the scenarios, we fix $\kappa = 10^{-2}$. We recall that the parameter $\bar{\epsilon}$ controls the external voltage, whereas $\dot{\varphi}_\Delta (0)$ tunes the initial voltage of the island.

\begin{figure*}[htb!]
    \captionsetup{justification=raggedright}
    \centering
    \includegraphics[width=15.0 cm]{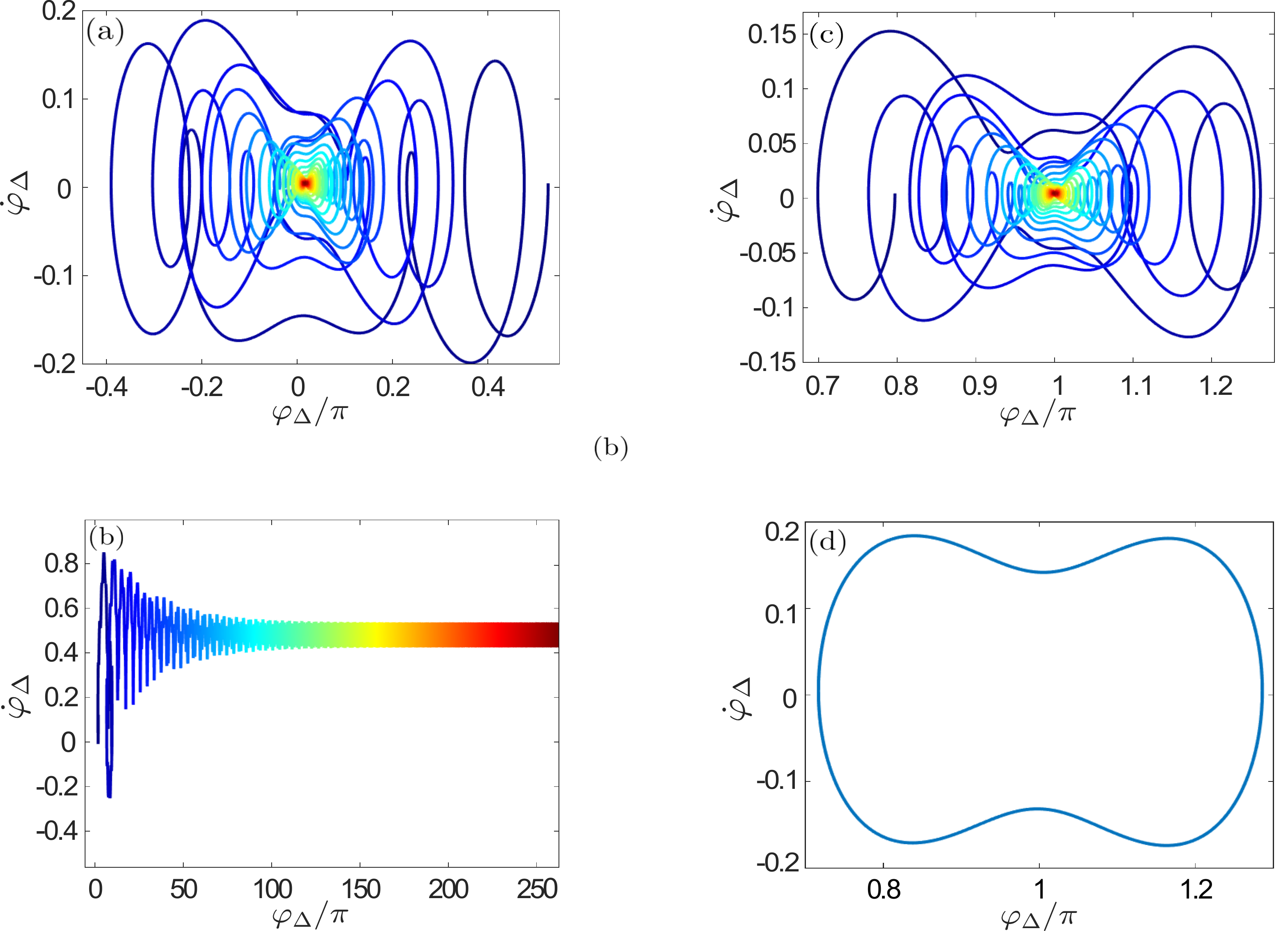}
    \caption{Trajectories in the phase space of (a) $0$-stable, (b) unstable, (c) $\pi$-stable and (d) limit-cycle solutions, obtained for $\dot{\varphi}_\Delta (0) = 0$. For stable solutions, the phase exhibits damped oscillation with decreasing amplitude. Their trajectories converge to their respective fixed points. The change in colour of plots (a)-(c) from dark blue to black indicates the direction of time. (a) $0$-stable solution for $\bar{\epsilon} = 0.27$ and $\varphi_\Delta(0) = 0.5\pi$. (b) Unstable solution for $\bar{\epsilon} = 0.58$ and $\varphi_\Delta(0) = 0. 8\pi$. Here, the trajectory does not remain in a neighbourhood of one of the fixed points. In the mechanical equivalent description, it is like the pendulum continues to constantly rotate with an angle exceeding the range $\left[0,2\pi\right]$. (c) $\pi$-stable solution for $\bar{\epsilon} = 0.27$ and $\varphi_\Delta(0) = 0.8\pi$. (d) Limit-cycle solution ($1$-cycle) for $\bar{\epsilon} = 0.5$ and $\varphi_\Delta (0) = 0.7\pi$.}
    \label{figure:stable}
\end{figure*}
%
\begin{figure*}[htb!]
\captionsetup{justification=raggedright}
\centering
    \includegraphics[width=15.0 cm]{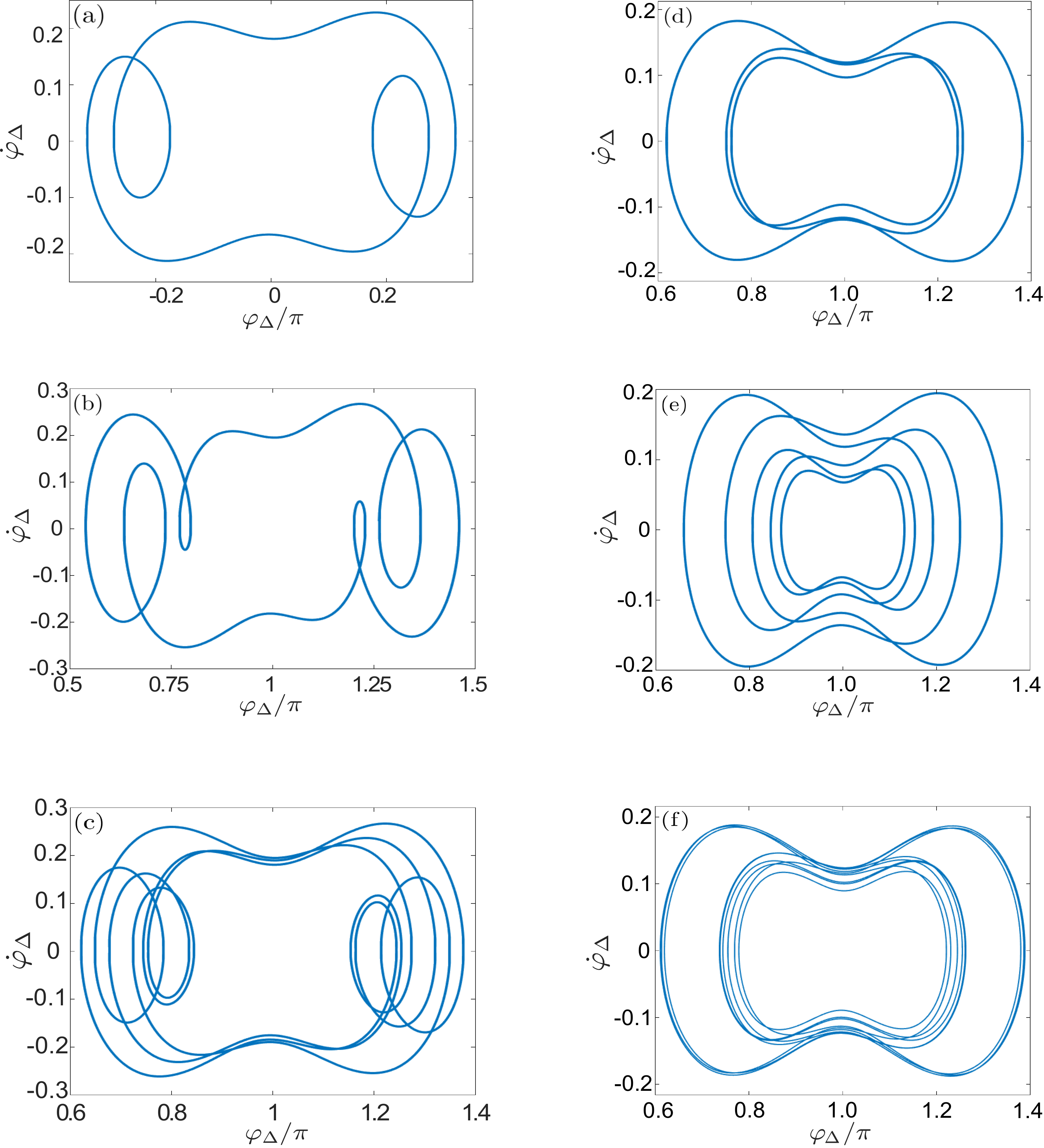}
    \caption{Trajectories of $3$-cycles, $5$-cycles and $9$-cycles appearing in the system dynamics with turning number larger than winding number (a)-(b)-(c) and with turning number equal to the winding number (d)-(e)-(f), obtained for $\dot{\varphi}_\Delta (0) = 0$. The $3$-cycles are obtained for (a) $\bar{\epsilon} = 0.37$, $\varphi_\Delta (0) = 0.54\pi$ and for (d) $\bar{\epsilon} = 0.52769$, $\varphi_\Delta (0) = 0.94458\pi$. The $5$-cycle are obtained for (b) $\bar{\epsilon} = 0.318828$, $\varphi_\Delta (0) = 0.727783\pi$ and (e) $\bar{\epsilon} = 0.50061$, $\varphi_\Delta (0) = 0.939941\pi$. The $9$-cycles are obtained for (c) $\bar{\epsilon} = 0.385088$, $\varphi_\Delta (0) = 0.827393\pi$ and (f) $\bar{\epsilon} =0.529131$, $\varphi_\Delta (0) = 0.942627\pi$.}
    \label{merged}
\end{figure*}
%

We numerically determine the asymptotic behaviour of the solutions $\varphi_\Delta(t)$, identifying four general types of attractors as a function of the dimensionless driving amplitude $\bar{\epsilon}$ and initial conditions $\left[\varphi_\Delta(0), \dot{\varphi}_\Delta(0)\right]$: $0$-stable, $\pi$-stable, unstable, and limit-cycle solutions. The appearance of four different types of attractors was previously discussed in the literature~\cite{acheson1995multiple,butikov2005complicated,bartuccelli2001dynamics}, where the rotor dynamics is analyzed as a function of the initial conditions and the system parameters: driving amplitude, gravity and, in some cases, dissipation. In our analysis, the solutions 0-stable and $\pi$-stable are such that $\varphi_\Delta=0$ and $\varphi_\Delta=\pi$ are stable fixed points (see figures~\ref{figure:stable}a and~\ref{figure:stable}c). The latter is reminiscent of the stabilization of the inverted position in the Kapitza pendulum in the presence of gravity. Unstable solutions correspond to asymptotically rotating solutions~\cite{xu2005rotating}, which are not characterized here any further (see figure~\ref{figure:stable}b). 
%
%
%
\begin{figure*}[htb!]
\captionsetup{justification=raggedright}
\centering
    \includegraphics[width=15.0cm]{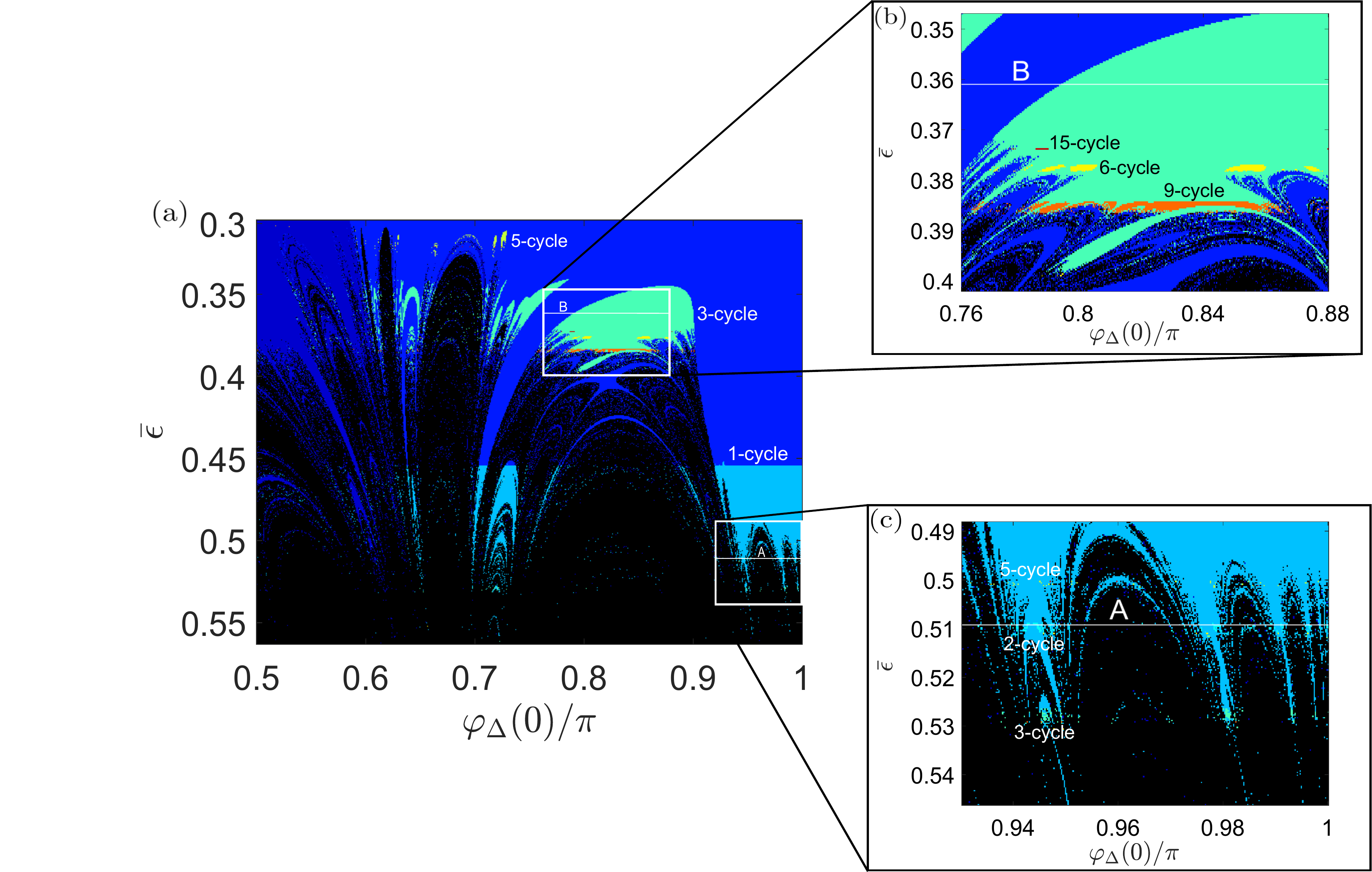}
    \caption{(a) Stability diagram for a parameters space comprised of $896\times2048$ points of $\varphi_{\Delta} \in [\frac{\pi}{2},\pi]$ in the horizontal axis, $\bar{\epsilon}\in[0.3,0.56]$ in the vertical axis and $\dot{\varphi}_\Delta (0) = 0$. The diagram is divided into four regions by the main attractors, each labelled with a different colour: black stands for unstable, dark blue for $0$-stable, blue for $\pi$-stable and the remaining colours represent the different n-cycles, which combine to form the limit-cycles region. The most likely are $1$-cycle (light blue), $3$-cycle (green) and $5$-cycle (light green), divided into three different subregions. However, the limit-cycle regions exhibit stripes and domains exhibiting higher-order n-cycles. (b) Inset of (a), where a region of $2$-cycles (cyan) appears along line A, standing for $\epsilon = 0.509253$. (c) Inset of (a), exhibiting subregions of $6$-cycles (yellow), $9$-cycles (orange) and a narrow stripe of $15$-cycles (red) for $\bar{\epsilon} = 0.373$ . Line B is for $\bar{\epsilon} = 0.361176755$.}
    \label{cycle}%
\end{figure*} 
%

The limit-cycle solutions are periodic orbits around $0$ and $\pi$ that exhibit a varying degree of complexity, depending on the driving amplitude and the initial conditions. In our work, we characterize these trajectories in terms of their number of intersections with the $\dot{\varphi}_\Delta = 0$ axis in the phase space (see figures~\ref{figure:stable}d and~\ref{merged}): if a periodic trajectory crosses the $\dot{\varphi}_\Delta=0$ axis 2n times in a period, it is classified as an n-cycle (see~\ref{app3}). It is straightforward to realize that the number of intersections n coincides with the number of loops, which is characterized by the turning number $t$. In addition, for a closed trajectory, it is possible to define the winding number $w$, counting the number of rotations around a given point $\vec{O}$.  Choosing in our case $\vec{O} = \left( \pi,0 \right)$ or $\left( 0,0 \right)$, we find both n-cycles for which $w = t$ (figure~\ref{merged}d-\ref{merged}e-\ref{merged}f) and cycles for which $w < t$ (figure~\ref{merged}a-\ref{merged}b-\ref{merged}c): trajectories execute loops on one side of the phase space before passing to the other side. We note that the same n-cycles can occur both with $t = w$  and $t > w$. In the literature, limit-cycle solutions have been classified in different ways (see~\cite{acheson1995multiple} and~\cite{butikov2002subharmonic}), but, to our knowledge, their topological properties have not been considered. In~\cite{acheson1995multiple}, these trajectories are described in terms of their multiple nodding behaviour observed in terms of the driving amplitude, dissipation and gravity, where a nodding corresponds to an instance where the pendulum has reached its local extremum away from the vertical position and starts to move back towards it: $n_A$ refers to the number of noddings counted in one-half of a limit-cycle. Comparing the limit-cycles analyzed in~\cite{acheson1995multiple} with our topological description, we find that in~\cite{acheson1995multiple} only multiple nodding trajectories with $w = 1$ are considered and that 
$n = 2n_A - w$. In~\cite{butikov2002subharmonic}, where zero-gravity case is considered, the limit-cycle solutions are described in terms of the subharmonic resonance, a phenomenon occurring when the driving frequency of the pivot is an integer multiple of the frequency of the cyclic trajectory of the pendulum. However, a clear connection between a specific subharmonic and a trajectory with certain number of nods is not drawn.
%
%

We have extensively analyzed the case in which $\bar{\epsilon}$ and $\varphi_\Delta (0)$ are varied and we fix $\dot{\varphi}_\Delta (0) = 0$, i.e. the JJs share the same initial voltage drop $V_\mathrm{g} / 2$. The results are summarized in the stability diagram of figure~\ref{cycle}, where the distribution of the attractors and n-cycles in the parameters space is reported. The main attractors ---unstable, $0$-stable, $\pi$-stable and limit-cycles (composed of all the n-cycles together)--- form four different regions in the stability diagram, showing a fractal geometry and high sensitivity to initial conditions (see figure~\ref{cycle}): the same behaviour is also encountered for the n-cycles subregions. From figure~\ref{cycle}a it is possible to see that $1$-cycle, $3$-cycle and $5$-cycle solutions identify three disjoint subsets, still exhibiting intricate geometries. A closer examination of the $1$-cycle (figure~\ref{cycle}c) and  $3$-cycle (figure~\ref{cycle}b) regions displays a richer scenario where higher-order n-cycles appear. For instance, in the 1-cycle region, a 2-cycle stripe appears along $\bar{\epsilon} \simeq 0.509253$ (line A in figure~\ref{cycle}c), along with (less prominent) 5-cycle and 3-cycle stripes for $\bar{\epsilon}\simeq 0.5$ and $\bar{\epsilon}\simeq 0.53$ (see figure~\ref{cycle}c). In the 3-cycle region (see figure~\ref{cycle}b), respectively around $\bar{\epsilon}\simeq 0.385$ and $\bar{\epsilon}\simeq 0.377$, 9-cycle and 6-cycle stripes appear, in addition to a (faint) 15-cycle stripe for $\bar{\epsilon} \simeq 0.373$ and a narrow line of 18-cycles below the 9-cycle stripe (for a detailed description of the attractors and the n-cycles analysis, see~\ref{app3}).
\begin{table}[htb!]
\caption{List of the n-cycles encountered in the stability diagram of figure~\ref{cycle} with their absolute, relative frequency and fractal dimension.}
\centering
\begin{tabular}{|c|c|c|c|}
\cline{1-4}
\multicolumn{1}{|c|}{\begin{tabular}[c]{@{}c@{}}$n$\\cycle\end{tabular}} & \begin{tabular}[c]{@{}c@{}}Relative\\ frequency\end{tabular} & \begin{tabular}[c]{@{}c@{}}Absolute\\ frequency\end{tabular} & \begin{tabular}[c]{@{}c@{}}Fractal \\ dimension\end{tabular} \\ \cline{1-4}
1                                                                               & 49.50                                                       & 73731   & 1.144   \\ \cline{1-4}
2                                                                               & 0.20                                                        & 294  &   0.875   \\ \cline{1-4}
3                                                                               & 47.70                                                       & 71118 &   1.310 \\ \cline{1-4}
4                                                                               & 0.01                                                        & 14  &    0.315 \\ \cline{1-4}
5                                                                              & 0.89                                                        & 1334&   1.021 \\ \cline{1-4}
6                                                                              & 0.43                                                        & 643 &     0.884 \\ \cline{1-4}
7                                                                              & 0.01                                                        & 16  &    0.354 \\ \cline{1-4}
9                                                                              & 1.16                                                         & 1772 &   0.988 \\ \cline{1-4}
15                                                                              & 0.05                                                        & 73 &     0.580  \\ \cline{1-4}
18                                                                              & 0.05                                                        & 75  &   0.482 \\ \cline{1-4}
\end{tabular}
\label{nods}
\end{table}

The attractors in the stability diagram exhibit an intricate distribution in terms of the parameters and high sensitivity to initial conditions. Our numerical investigation provides further insight into these properties of the stability diagram: we characterize the fractal dimension of each region in the stability diagram as a function of the two parameters $\bar{\epsilon}$ and $\varphi_\Delta(0)$. We do this by estimating the Hausdorff dimension of their borders in figure~\ref{cycle}a with the box-counting method~\cite{stewart2001towards}: the unstable solutions have a Hausdorff dimension $\delta_H \simeq 1.520$, $0$-stable $\delta_H \simeq 1.327$, $\pi$-stable $\delta_H \simeq 1.307$ and limit-cycles (comprised of all the n-cycles) $\delta_H \simeq 1.201$. The list of different n-cycles observed in our simulations is reported in table~\ref{nods}, along with the fractal dimensions of the regions they form in the stability diagram in figure~\ref{cycle} (for a detailed discussion about the estimation of the Hausdorff dimensions, see~\ref{app3}).
%
\begin{figure*}[htb!]
\captionsetup{justification=raggedright}
\centering
    \includegraphics[width=15.0cm]{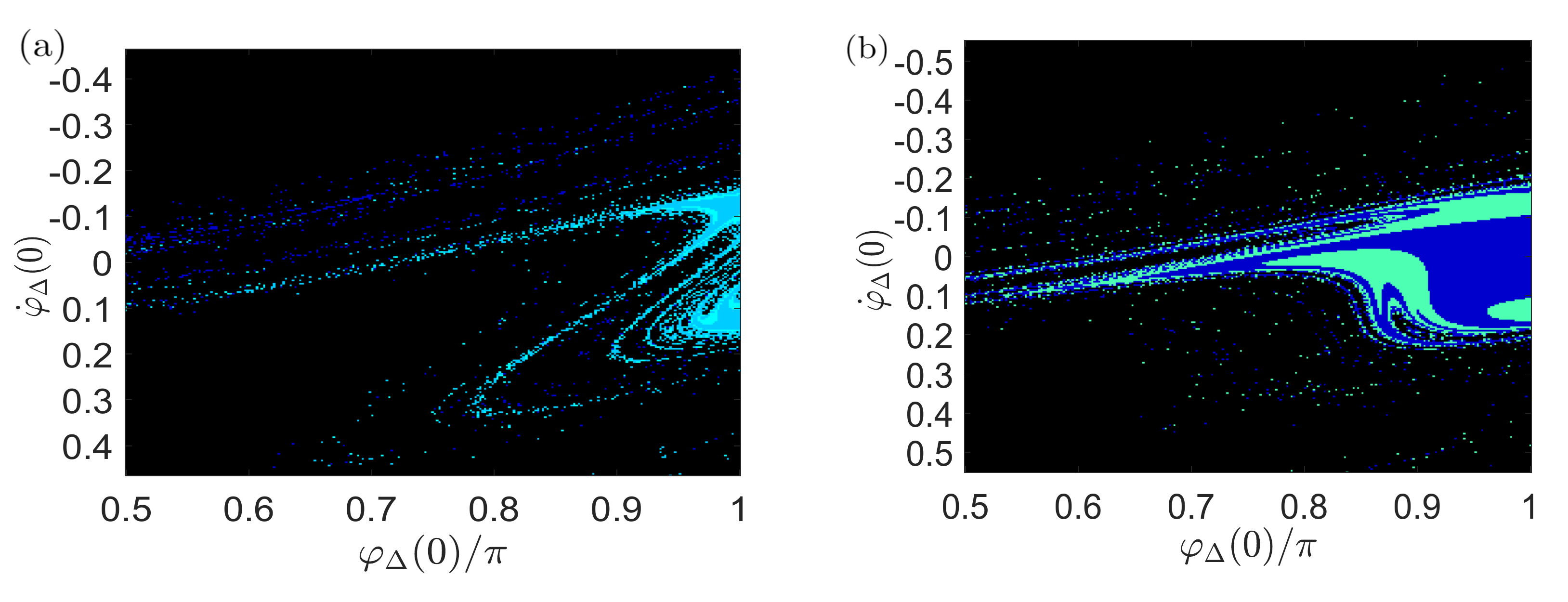}
    \caption{Stability diagrams for the parameters space $\left[\varphi_\Delta (0), \dot{\varphi}_\Delta (0) \ne 0\right]$ and for fixed values of $\bar{\epsilon}$: (a) for $\bar{\epsilon} = 0.509253$ (line A in figure~\ref{cycle}c) and (b) for $\bar{\epsilon}=0.361176755$ (line B in figure~\ref{cycle}b). The colour-coding is the same as figure~\ref{cycle}.}
    \label{Fig5}%
\end{figure*} 
%

The case $\dot{\varphi}_\Delta(0) \ne 0$, i.e. asymmetric initial voltage drop $V_\mathrm{g} / 2 \mp \dot{\phi}_\Delta (0)$ across the two JJs, has been analyzed for $\bar{\epsilon}\simeq 0.509253$ (line A in figure~\ref{cycle}c) and $\bar{\epsilon} \simeq 0.361176755$ (line B in figure~\ref{cycle}b) varying both the initial conditions $\left[\varphi_\Delta(0), \dot{\varphi}_\Delta (0)\right]$. The attractors appearing from these parameter settings are of the same type as the case $\dot{\varphi}_\Delta(0) = 0$: the results of the numerical analysis are shown in figure~\ref{Fig5}, where we report the basins of attraction related to the different attractors. In addition, we note that the number n of the intersection of the n-cycles seem not to change when we move away from the condition $\dot{\varphi}_\Delta (0) = 0$, at least for the lines A and B: comparing figures~\ref{cycle}c and~\ref{Fig5}a, line A exhibits $1$- and $2$-cycles both for $\dot{\varphi}_\Delta (0) = 0$ and $\dot{\varphi}_\Delta (0) \ne 0$ and the same happens for line B, which exhibits only $1$-cycles throughout its dynamics (see figures~\ref{cycle}b and~\ref{Fig5}b).

In~\cite{bartuccelli2001dynamics}, the authors show the basins of attraction of the rotor dynamics obtained as a function of the initial conditions and for fixed values of the driving amplitude and the gravity. Even though we use a different parameters space, our basins of attraction in figure~\ref{Fig5} exhibit a fractal structure similar to the one reported in~\cite{bartuccelli2001dynamics}: here, we emphasize this property by computing the Hausdorff dimension of the borders of the attractors, in the same way as the case $\dot{\varphi}_\Delta(0) = 0$. For the case depicted in figure~\ref{Fig5}a the unstable solutions have dimension $\delta_H = 1.233$, $\delta_H = 1.036$ for $0$-stable, $\delta_H = 1.250$ for $1$-cycles and $\delta_H = 1.123$ for $2$-cycles, whereas for figure~\ref{Fig5}b the Hausdorff dimensions are $\delta_H = 1.434$  for unstable solutions, $\delta_H = 1.414$ for $0$-stable, $\delta_H = 1.310$ for $\pi$-stable and $\delta_H = 1.246$ for $3$-cycle.

\section{Conclusions}
Through a classical mechanics approach that takes into account voltage sources as semi-holonomic constraints, we have shown that the semiclassical dynamics of a circuit comprising two Josephson junctions in series can be described by the same dynamical equations of the planar (zero-gravity) parametrically driven pendulum. In our numerical analysis, we identify four different types of attractors ---$0$-and $\pi$-stable, unstable and limit-cycle--- accessible by tuning the dynamical parameters of the system. Moreover, we classify the stable limit-cycle solutions as n-cycles, where n equals the turning number of the limit-cycle. In addition, we describe the distribution of the attractors in the stability diagram and basins of attraction, which reveals the chaotic behaviour of the system dynamics by characterizing the different attractors according to their Hausdorff dimension.

\section*{Acknowledgement}
FM and JM acknowledge financial support from the Research Council of Norway (Grant No. 333937) through participation in the QuantERA ERA-NET Cofund in Quantum Technologies (project MQSens) implemented within the European Union’s Horizon 2020 Programme. The computations were performed on resources provided by Sigma2 - the National Infrastructure for High Performance Computing and
Data Storage in Norway.

\appendix
\section{\label{app1}Deriving the Lagrangian and the Hamiltonian}
We formulate the Lagrangian in terms of the fluxes $\phi_i (t)$ and voltages $\dot{\phi}_i (t) = V_i (t)$  associated with the nodes $i = g,I,o$, where the flux is taken as a coordinate-like variable.  Due to the grounding of the circuit, it is implied that $V_{\rm o} = \dot{\phi}_\mathrm{o} = 0$ but we do not explicitly enforce this constraint until later on in the derivation of the equations of motion to keep the calculation as general as possible.

The kinetic term of the Lagrangian takes the form
\begin{equation}
    \mathcal{T} = \frac{C_{\mathrm{J}1}}{2} \left( \dot{\phi}_\mathrm{g} - \dot{\phi}_\mathrm{I} \right)^2 
    + \frac{C_{\mathrm{J}2}}{2} \left( \dot{\phi}_\mathrm{I} - \dot{\phi}_\mathrm{o} \right)^2
\end{equation}
and the potential energy part arising from the Josephson junctions
\begin{equation}
\begin{split}
    \mathcal{U} &= - E_{\mathrm{J}1}\cos\left(\frac{2\pi}{\Phi_0}(\phi_\mathrm{g} - \phi_\mathrm{I})\right) -       E_{\mathrm{J}2}\cos\left(\frac{2\pi}{\Phi_0} (\phi_\mathrm{I} - \phi_\mathrm{o}) \right) \\
    &= - E_{\mathrm{J}1}\cos\left(\varphi_\mathrm{g} - \varphi_\mathrm{I}\right) -       E_{\mathrm{J}2}\cos\left(\varphi_\mathrm{I} - \varphi_\mathrm{o}) \right), 
\end{split}
\end{equation}
where the phase-flux relation $\varphi_\mathrm{i} =2\pi\phi_\mathrm{i} /\Phi_0$ is used  and $\Phi_0 = h/(2e)$ is the flux quantum. Together, they form the Lagrangian $\mathcal{L}_0 = \mathcal{T} - \mathcal{U}$. To take into account the external voltage source in the analysis of the system, we consider it as a constraint on the variables of the system ($\dot{\phi}_\mathrm{g} - \dot{\phi}_\mathrm{o} - V_\mathrm{g} = 0$) and we enforce it in the Lagrangian by using the undetermined Lagrange multiplier method, which requires introducing the Lagrangian (charge) multiplier $Q$.  Note that the constraint is semi-holonomic, linear in the derivatives of the generalized coordinates, which allows us to obtain the same equations of motion as from the holonomic one $\phi_\mathrm{g} - \phi_\mathrm{o} - V_\mathrm{g} t = 0$ with the Lagrange multiplier $I$ related to $Q$ by $I= -\dot{Q}$. This equivalence is not true in general for arbitrary constraints on the coordinate derivates, but it can be shown for the linear case by considering the action 
\begin{equation}
    \mathcal{S}_1 = \int_{t_1}^{t_2} dt \left[\mathcal{L}_0 + I(t)\left(\phi_\mathrm{g} - \phi_\mathrm{o} - V_\mathrm{g} t\right) \right]= \mathcal{S}_{0} + \int_{t_1}^{t_2} dt \left[I(t)(\phi_\mathrm{g} - \phi_\mathrm{o} - V_\mathrm{g} t)\right],
\end{equation}
related to the holonomic constraint with multiplier $I$ and the action
\begin{equation}
    \mathcal{S}_2 = \int_{t_1}^{t_2} dt \left[\mathcal{L}_0 + Q(t)\left(\dot{\phi}_\mathrm{g} - \dot{\phi}_\mathrm{o} - V_\mathrm{g}\right) \right] = \mathcal{S}_0 + \int_{t_1}^{t_2} dt \left[Q(t)(\dot{\phi}_\mathrm{g} - \dot{\phi}_\mathrm{o} - V_\mathrm{g}) \right]
\end{equation}
related to the semi-holonomic constraint with multiplier $Q$. The action $\mathcal{S}_2$ can be written as
\begin{equation}
\begin{split}
    \mathcal{S}_2 &= \mathcal{S}_0 - \int_{t_1}^{t_2} dt \left[\dot{Q}(t)\left(\phi_\mathrm{g} - \phi_\mathrm{o} - V_\mathrm{g} t\right) \right] + \left[Q(t)\left(\phi_\mathrm{g} - \phi_\mathrm{o} - V_\mathrm{g} t \right)\right]_{t_1}^{t_2} \\ &= \mathcal{S}_0 - \int_{t_1}^{t_2} dt \left[\dot{Q}(t)\left(\phi_\mathrm{g} - \phi_\mathrm{o} - V_\mathrm{g} t\right) \right] + const.
\end{split}
\end{equation}
Thus, by taking $I = - \dot{Q}$, the actions $\mathcal{S}_1$  and $\mathcal{S}_2$ differ by a constant, i.e. they are minimized by the same trajectory and thus lead to the same equations of motion. 

With this scheme in mind, we write the constrained Lagrangian
\begin{equation}
    \mathcal{L} = \mathcal{T} - \mathcal{U} 
    + Q \left( \dot{\phi}_\mathrm{g} - \dot{\phi}_\mathrm{o} - V_\mathrm{g} \right).
\end{equation}
The node charges that are the conjugate momenta to the fluxes can be straightforwardly calculated
\begin{subequations}
\begin{align}
    Q_\mathrm{I} &= \frac{\partial \mathcal{L}}{\partial \dot{\phi}_\mathrm{I}} 
        = C_{\mathrm{J}1} \left( \dot{\phi}_\mathrm{I} - \dot{\phi}_\mathrm{g} \right) 
        + C_{\mathrm{J}2} \left( \dot{\phi}_\mathrm{I} - \dot{\phi}_\mathrm{o} \right) , \\
    Q_\mathrm{g} &= \frac{\partial \mathcal{L}}{\partial \dot{\phi}_\mathrm{g}} 
        = C_{\mathrm{J}1} \left( \dot{\phi}_\mathrm{g} - \dot{\phi}_\mathrm{I} \right) + Q , \\
    Q_\mathrm{o} &= \frac{\partial \mathcal{L}}{\partial \dot{\phi}_\mathrm{o}} 
        = C_{\mathrm{J}2} \left( \dot{\phi}_\mathrm{I} - \dot{\phi}_\mathrm{o} \right) - Q ,
\end{align}
\end{subequations}
leading to the Hamiltonian $\mathcal{H} = \mathcal{H}(\vec{Q},\vec{\phi})$ 
given by the Legendre transformation of the Lagrangian
\begin{equation}
\begin{split}
    \mathcal{H} =& \sum_{i} Q_i \dot{\phi}_i - \mathcal{L} \\
        =& 4 E_{\mathrm{C}1} \left(n_\mathrm{g} - N \right)^2
        + 4 E_{\mathrm{C}2} \left(n_\mathrm{I} + n_\mathrm{g} - N \right)^2 
        + 2 e N V_\mathrm{g} + \mathcal{U} 
\end{split}
\label{SI:hamiltonian}
\end{equation}
with $E_{\mathrm{C}i} = e^2/(2 C_{\mathrm{J}i})$
and the charges are given as the number of Cooper pairs $n_i = Q_i/(2e)$ 
with the Lagrange multiplier $N = Q/(2e)$.

\section{\label{app2}Deriving Hamilton's equations and the dynamical equation for superconductive phase}

The canonical conjugate variables in our formalism are the flux $\phi_i(t)$ and the charge $Q_i(t)$ at each node $i = g,I$. However, in this case, we write the Hamiltonian in terms of the number of Cooper pairs $n_i = Q_i/(2e)$ and the phase $\varphi_i = 2\pi\phi_i/\Phi_0$. For this reason, taking $f$ and $g$ as two generic functions of the canonical variables, the usual Poisson brackets become
\begin{equation}
\{f,g\} = \sum_{i = g,I} \frac{\partial f}{\partial \phi_i}\frac{\partial g}{\partial Q_i} - \frac{\partial f}{\partial Q_i}\frac{\partial g}{\partial \phi_i} = \frac{1}{\hbar} \sum_{i = g,I} \frac{\partial f}{\partial \varphi_i}\frac{\partial g}{\partial n_i} - \frac{\partial f}{\partial n_i}\frac{\partial g}{\partial \varphi_i},
\end{equation}
from which we derive Hamilton's equations for the variables at nodes $i = g, I$
\begin{subequations}
\begin{align}
    \dot{n}_i &= \{n_i,\mathcal{H}\} = -\frac{1}{\hbar}\frac{\partial \mathcal{H}}{\partial \varphi_i}, \\
    \dot{\varphi}_i &= \{\varphi_i, \mathcal{H}\} = \frac{1}{\hbar}\frac{\partial \mathcal{H}}{\partial n_i}.
\end{align}
\end{subequations}
Explicitly, Hamilton's equations are
\begin{subequations}
\begin{align}
    \dot{n}_\mathrm{I} &=  -\frac{1}{\hbar} \left[ E_{\mathrm{J}1} \sin(\varphi_\mathrm{I} - \varphi_\mathrm{g})
            + E_{\mathrm{J}2} \sin(\varphi_\mathrm{I} - \varphi_\mathrm{o}) \right], \label{SI:nI} \\
    \dot{n}_\mathrm{g} &= \frac{1}{\hbar} E_{\mathrm{J}1} \sin(\varphi_\mathrm{I} - \varphi_\mathrm{g}), \label{SI:nG} \\
    \dot{\varphi}_\mathrm{I} &= \frac{8}{\hbar} E_{\mathrm{C}2} (n_\mathrm{I} + n_\mathrm{g} - N), \label{SI:phiI} \\
    \dot{\varphi}_\mathrm{g} &= \frac{8}{\hbar} \left[ E_{\mathrm{C}2} n_\mathrm{I} 
            + \left( E_{\mathrm{C}1} + E_{\mathrm{C}2} \right) 
            \left( n_\mathrm{g} - N \right) \right] . \label{SI:phiG}
\end{align}
\end{subequations}

The voltage constraint $\omega_\mathrm{g} = \dot{\varphi}_\mathrm{g} - \dot{\varphi}_\mathrm{o} = \dot{\varphi}_\mathrm{g}$  gives that the time evolution of the superconducting phase  (related to the voltage) on the lead $g$ given by equation~\eqref{SI:phiG} is a constant  $\omega_\mathrm{g} = 2 \pi V_\mathrm{g}/\Phi_0$ and we can solve for the Lagrange multiplier
\begin{equation}
\begin{split}
    N &= \frac{E_{\mathrm{C}2}}{E_{\mathrm{C}1} + E_{\mathrm{C}2}} n_\mathrm{I} + n_\mathrm{g}
        -\frac{\hbar}{8} \frac{\omega_\mathrm{g}}{E_{\mathrm{C}1} + E_{\mathrm{C}2}} \\
        &= \frac{C_{\mathrm{J}1}}{C_{\mathrm{J}1} + C_{\mathrm{J}2}}n_\mathrm{I} + n_\mathrm{g} - \frac{1}{2e}\frac{C_{\mathrm{J}1}C_{\mathrm{J}2}}{C_{\mathrm{J}1} + C_{\mathrm{J}2}}V_\mathrm{g}.
\end{split}
\label{SI:Lagrangian_multiplier}
\end{equation}
The physical interpretation of the multiplier $N$ can be given by considering a slightly different setting of the circuit, in which the tunnelling across junctions is neglected and the external voltage source is replaced by the capacitor $C_{\mathbin{\|}} = C_{\mathrm{J}1}C_{\mathrm{J}2} / \left(C_{\mathrm{J}1} + C_{\mathrm{J}2}\right)$ with a voltage drop $V_\mathrm{g}$: the last term in the expression of $N$ represents then the amount of Cooper pairs stored on $C_{\mathbin{\|}}$. We can write the number of Cooper pairs on $\mathrm{g}$ and $\mathrm{I}$ in terms of the numbers of Cooper pairs $n_1$ and $n_2$ on the two JJ capacitances as $n_{\mathrm{g}} = n_1 + \frac{C_{\mathbin{\|}}V_\mathrm{g}}{2e}$ and $n_\mathrm{I} = n_2 - n_1$. Substituting these expressions in equation~\ref{SI:Lagrangian_multiplier}, we obtain
\begin{equation}
    N = C_{\mathbin{\|}} \left(\frac{n_1}{C_{\mathrm{J}1}} + \frac{n_2}{C_{\mathrm{J}2}} \right) = \frac{1}{2e} C_{\mathbin{\|}}\left( V_1 + V_2 \right) = \frac{1}{2e} C_{\mathbin{\|}} V_\mathrm{g}.
\end{equation}
Thus, the Lagrange multiplier $N$ corresponds to the number of Cooper pairs on the capacitance $C_{\|}$, which, considering the original circuit with the voltage source, can be regarded as the charge stored by the external voltage $V_\mathrm{g}$ on the junctions capacitances $C_{\mathrm{J}1}$ and $C_{\mathrm{J}2}$ in series.  Furthermore, we note that the solution of the Lagrange multiplier allows us to write the Hamiltonian in equation \eqref{SI:hamiltonian} in a more familiar form 
\begin{equation}
    \mathcal{H} = 4 E_\mathrm{C} \left( n_\mathrm{I} - \bar{n}_\mathrm{g} \right)^2
        - \frac{C_{\mathrm{J}1}}{2} V_\mathrm{g}^2  
        + 2 e n_\mathrm{g} V_\mathrm{g} 
        + \mathcal{U} ,
\label{SI:hamiltonian2}
\end{equation}
where $\bar{n}_\mathrm{g} = - C_{\mathrm{J}1} V_\mathrm{g}/(2e)$ is the conventional gate charge divided by the Cooper pair unit of charge.

Hamilton's equations~\eqref{SI:phiI} and~\eqref{SI:phiG} can now be rewritten as
\begin{subequations}
\begin{align}
    \dot{\varphi}_\mathrm{I} &= \frac{8}{\hbar} \frac{E_{\mathrm{C}1} E_{\mathrm{C}2}}{E_{\mathrm{C}1} + E_{\mathrm{C}2}} n_\mathrm{I} + \frac{E_{\mathrm{C}2}}{E_{\mathrm{C}1} + E_{\mathrm{C}2}} \omega_\mathrm{g}, \label{SI:phiI2} \\
    \dot{\varphi}_\mathrm{g} &= \omega_\mathrm{g} , \label{SI:phiG2}
\end{align}
\end{subequations}
the latter implying $\varphi_\mathrm{g} (t) = \varphi_\mathrm{g} (0) + \omega_g t$.

Differentiating equation~\eqref{SI:phiI2} with respect to time and plugging in equation~\eqref{SI:nI}, we obtain
\begin{equation}
\begin{split}
    \ddot{\varphi}_\mathrm{I} &= \frac{8}{\hbar} \frac{E_{\mathrm{C}1} E_{\mathrm{C}2}}{E_{\mathrm{C}1} + E_{\mathrm{C}2}} \dot{n}_\mathrm{I} \\
        &=- \frac{8 E_\mathrm{C}}{\hbar^2}\left[ E_{\mathrm{J}1}\sin(\varphi_\mathrm{I} - \varphi_\mathrm{g}) + E_\mathrm{J2}\sin(\varphi_\mathrm{I})\right]
\end{split}
\end{equation}
with $E_\mathrm{C} = e^2/(2(C_{\mathrm{J}1} + C_{\mathrm{J}2}))$.
If we now consider identical junctions, i.e. $E_{\mathrm{J}1} = E_{\mathrm{J}2} = E_{\mathrm{J}}$
\begin{equation}
    \ddot{\varphi}_\mathrm{I} = - \frac{8E_\mathrm{C} E_\mathrm{J}}{\hbar^2}\left(\sin{\varphi_\mathrm{I}} - \sin{(\varphi_\mathrm{g} - \varphi_\mathrm{I})}\right),
\end{equation}
and exploiting $\sin\left(A\right) - \sin\left(B\right) = 2\cos\left(\frac{A+B}{2}\right)\sin\left(\frac{A-B}{2}\right)$ we end up with the second order differential equation for the phase $\varphi_{\mathrm{\Delta}} = \varphi_{\mathrm{I}} - \varphi_{\mathrm{g}}/2$
\begin{equation}
    \ddot{\varphi}_\Delta = -2\Omega^2_{\mathrm{J}}\cos\left[\frac{\varphi_\mathrm{g} (t)}{2}\right]\sin{\varphi_\Delta},
    \label{kapitza3}
\end{equation}
where $\Omega_\mathrm{J} = \sqrt{8 E_\mathrm{C} E_\mathrm{J}}/\hbar$ is the Josephson frequency.

Taking the dimensionless variable $\tau = \omega_\mathrm{g} t/2$, the derivatives with respect to the time become $\partial_t = \omega_\mathrm{g} \partial_\tau/2$ and $\partial_t^2 = \omega_\mathrm{g}^2 \partial_\tau^2/4$. This change of variable in equation \eqref{kapitza3} gives us the dimensionless differential equation
\begin{equation}
    \partial_\tau^2 \varphi_{\Delta} = -\bar{\epsilon}\cos\left(\tau\right)\sin{\varphi_{\Delta}},
\end{equation}
where
\begin{equation}
    \bar{\epsilon} = 8 \frac{\Omega_{\mathrm{J}}^2}{\omega_\mathrm{g}^2} = \frac{64 E_\mathrm{C} E_\mathrm{J}}{(2e)^2 V_\mathrm{g}^2}.
\end{equation}
%
\section{Derivation of the dissipative equation of motion}
\label{sec:RCSJ}

Let us derive the dissipative equation of motion for the superconductive phase using
the resistively and capacitively shunted junction model, where a Josephson
junction is placed in parallel, as the name suggests, with a capacitor and a resistor to
approximate loss mechanisms in the junction. To this end, we utilize the two Josephson
relations
\begin{subequations}
\label{eq:Josephson_relations}
\begin{align}
    I &= I_\mathrm{C} \sin \delta, \label{eq:Jrel1} \\ 
    \dot{\delta} &= \frac{2 e V}{\hbar}, \label{eq:Jrel2}
\end{align}    
\end{subequations}
where $I_\mathrm{C}$ is the critical current, $\delta$ the superconducting phase difference
across the junction, and $V$ the voltage across the junction.

For both JJs in our circuit, the current through them is
\begin{equation}
\label{eq:RCSJ_current}
    I_j = I_\mathrm{C} \sin \delta_j + C_j \dot{V}_j + \frac{V_j}{R_j} ,
\end{equation}
$j \in \left\{g,I\right\}$. Using the second Josephson relation, equation \eqref{eq:Jrel2},
we can reformulate the equation for the current through the junction as
\begin{equation}
\label{eq:RCSJ_current2}
    I_j = I_\mathrm{C} \sin \delta_j + \frac{\hbar}{2e} C_j \ddot{\delta}
        + \frac{\hbar}{2 e R_j} \dot{\delta} .
\end{equation}
Let us write the phase differences across the junctions using the explicit node phases so that
$\delta_1 = \varphi_\mathrm{g} - \varphi_\mathrm{I}$ and $\delta_2 = \varphi_\mathrm{I} - \varphi_\mathrm{o}$. The two JJs are in
series, so the electric current through them must be equal
\begin{equation}
\label{eq:equal_currents}
    \eqalign{I_\mathrm{C} \sin \left( \varphi_\mathrm{g} - \varphi_\mathrm{I} \right)
        + \frac{\hbar}{2e} C_{\mathrm{J}1} \left( \ddot{\varphi}_\mathrm{g} - \ddot{\varphi}_\mathrm{I} \right)
        + \frac{\hbar}{2 e R_1} \left( \dot{\varphi}_\mathrm{g} - \dot{\varphi}_\mathrm{I} \right) \\
    = I_\mathrm{C} \sin \left( \varphi_\mathrm{I} - \varphi_\mathrm{o} \right)
        + \frac{\hbar}{2e} C_{\mathrm{J}2} \left( \ddot{\varphi}_\mathrm{I} - \ddot{\varphi}_\mathrm{o} \right)
        + \frac{\hbar}{2 e R_2} \left( \dot{\varphi}_\mathrm{I} - \dot{\varphi}_\mathrm{o} \right).}
\end{equation}

We note that node 3 is grounded, i.e. $\varphi_\mathrm{o} = \dot{\varphi}_\mathrm{o} = \ddot{\varphi}_\mathrm{o} = 0$,
and due to the voltage source in the circuit, the phase of node 2 satisfies
\begin{eqnarray}
\label{eq:phase2}
    \varphi_\mathrm{g} = \varphi_\mathrm{g} (0) + \frac{2\pi}{\phi_0} V_\mathrm{g} t, \\
    \dot{\varphi}_\mathrm{g} = \frac{2\pi}{\phi_0} V_\mathrm{g} = \omega_\mathrm{g}.
\end{eqnarray}
Assuming identical junctions with $R_1 = R_2 = R$ and recalling that the critical current
relates to the Josephson energy so that $I_\mathrm{C} = \frac{2\pi}{\phi_0} E_\mathrm{J}$,
we can solve equation \eqref{eq:equal_currents} to obtain
\begin{equation}
\label{eq:SI_dissipative_EOM}
\begin{split}
    \ddot{\varphi}_\mathrm{I} =& \frac{1}{\hbar^2} \frac{4 e^2}{C_{\mathrm{J}1} + C_{\mathrm{J}2}} E_\mathrm{J}
                            \left[ \sin \left( \varphi_\mathrm{I} - \varphi_\mathrm{g} \right) + \sin \left( \varphi_\mathrm{I} \right) \right] \\
                    &+ \frac{1}{R \left( C_{\mathrm{J}1} + C_{\mathrm{J}2} \right)}
                            \left( - 2 \dot{\varphi}_\mathrm{I} + \dot{\varphi}_\mathrm{g} \right) .
\end{split}
\end{equation}
Following the treatment of the EOM in Appendix \ref{app2}, we obtain
\begin{equation}
\label{eq:SI_dissipative_EOM2}
    \ddot{\varphi}_\Delta = -2 \Omega_\mathrm{J}^2 \sin \varphi_\Delta \cos \frac{\varphi_\mathrm{g}}{2}
                            - \frac{2}{R \left( C_{\mathrm{J}1} + C_{\mathrm{J}2} \right)}
                                \dot{\varphi}_\Delta ,
\end{equation}
where $\frac{2}{R \left( C_{\mathrm{J}1} + C_{\mathrm{J}2} \right)}$ is the dissipation rate.
Writing this in dimensionless form, once again following the conventions of \ref{app2}, we recover equation \eqref{kapitza2} of the main text
\begin{equation}
    \partial_\tau^2 \varphi_\mathrm{\Delta} = -\bar{\epsilon}\cos\left(\tau\right)\sin{\varphi_{\Delta}} -\kappa \partial_\tau \varphi_{\mathrm{\Delta}}
    \label{eq:SI_kapitza2}
\end{equation}
with $\kappa = 4/\left[\omega_\mathrm{g}R (C_{\mathrm{J}1}+C_{\mathrm{J}2})\right]$.

\section{\label{app3} Computational methods}

A code written in Julia environment is used for the numerical simulation of equation \eqref{eq:SI_kapitza2}, corresponding to equation \eqref{kapitza2} in the main text. The integration time is long for the dissipation to take place and eventually stabilize the solutions. Once the trajectory is obtained, the data related to the intermediate transient time are discarded, highlighting only the behaviour of the trajectory in the long time regime, where its specific attractor shows up.

The classification in terms of attractors of the solutions is based on the threshold values $\varphi_{th} = 10^{-2}$ and $\dot{\varphi}_{th} = 5\cdot10^{-3} s^{-1}$ for the phase and its derivative, through the following criteria:
\begin{itemize}
    \item $\pi$-stable : a solution for which $|\max{\varphi_\Delta} - \pi| < \varphi_{th} $ and $|\max{\dot{\varphi}_\Delta}|< \dot{\varphi}_{th}$ (see figure~\ref{figure:stable}c);
    \item $0$-stable: a solution for which $|\max{\varphi_\Delta} - 0| < \varphi_{th} $ and $|\max{\dot{\varphi}_\Delta}|< \dot{\varphi}_{th}$ (see figure~\ref{figure:stable}a);
    \item cycle limit: a solution for which $|\max{\varphi_\Delta} - \pi| < \frac{\pi}{2}$ (or $|\max{\varphi_\Delta}| < \frac{\pi}{2}$ for limit-cycles around the downward fixed point) and $|\max{\dot{\varphi}_\Delta}|> \dot{\varphi}_{th}$;
    \item unstable: a solution for which $|\max{\varphi_\Delta}| > 2\pi$ (see figure~\ref{figure:stable}b).
\end{itemize}

Our first result is a stability diagram having a parameters space composed of $1024$ points of $\varphi_\Delta (0) \in [\frac{\pi}{2}, \pi]$ and $1024$ points of $\bar{\epsilon} \in [0.01, 0.6]$ and keeping $\dot{\varphi}_\Delta (0) = 0$, obtained through an optimized computational method. In the first step, we have taken just the extreme values of the parameters in their interval of definition, having 4 couples of points $(\varphi_{\mathrm{\Delta}}, \bar{\epsilon})$ or, in other words, a parameters space of $2\times2$ classified solutions. In the second step, we bisect the interval of each parameter, generating a $3\times3$ space in which one-quarter of the simulations come from the previous $2\times2$ case, and the remaining ones are simulated. This bisection scheme has been repeated up to the final result of a map composed of $1024\times1024$ solutions. Again, the bisection scheme is implemented on the map's portion corresponding to the parameters $\bar{\epsilon} \in [0.3,0.58]$ and $\varphi_\Delta (0) \in [0.5\pi, \pi]$, reported in figure~\ref{cycle}a. Here, we compute the fractal dimension of the regions generated by the attractors using the box-counting method.

\begin{figure*}[htb!]
\captionsetup{justification=raggedright}
\centering
    \includegraphics[width= 15 cm]{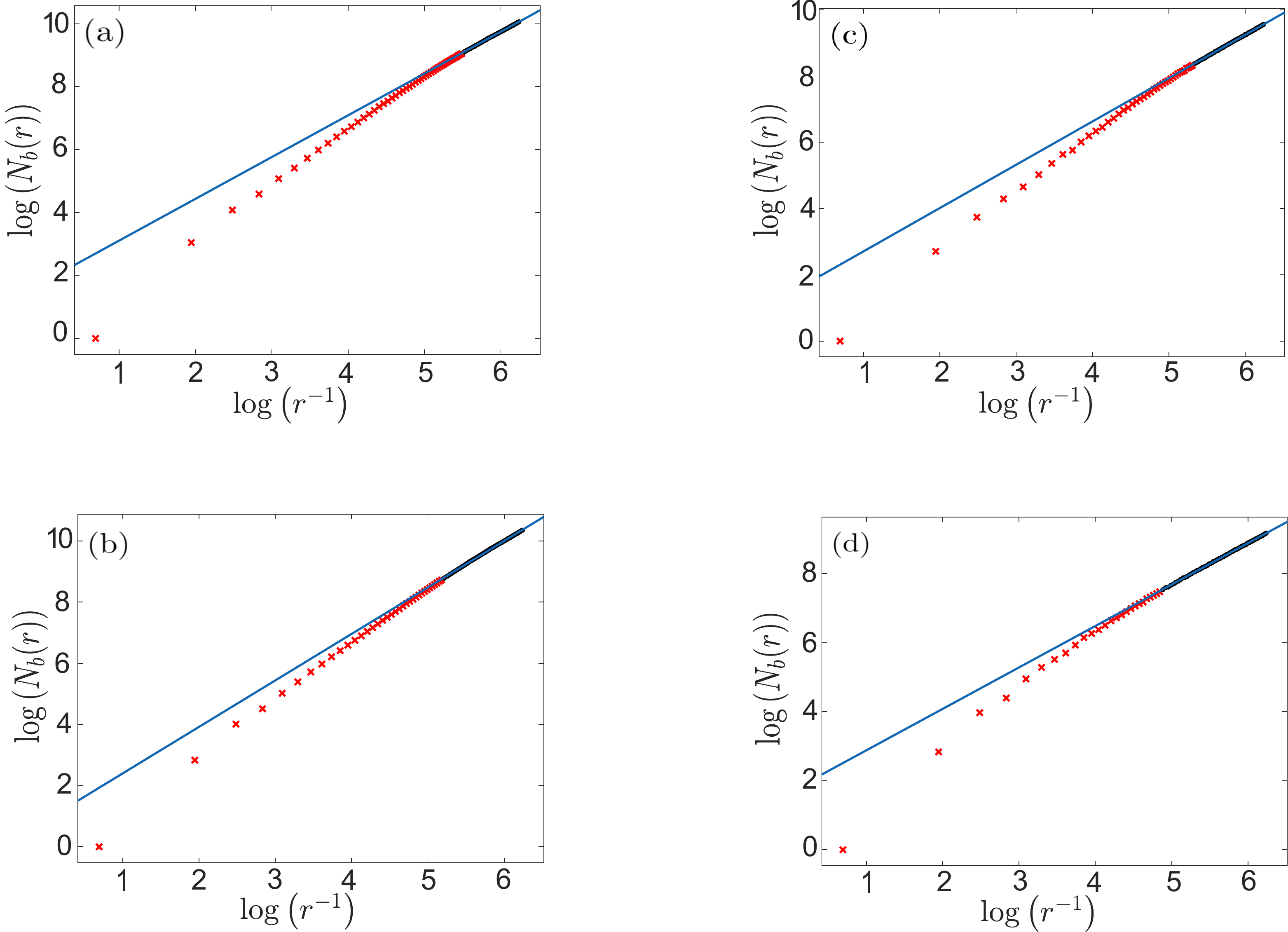}
    \caption{Estimation of the fractal dimensions $\delta_{\rm H}$ via linear best fits (blue
      lines) of the four different regions in figure~\ref{cycle}a. The different
      linear fits correspond to (a) $0-$stable solutions ($d_{\rm H} \simeq 1.327$), (b) unstable solutions ($d_{\rm H} \simeq 1.520$), (c) $\pi-$stable solutions ($d_{\rm H} \simeq 1.307$), (d)
      limit-cycle solutions, comprised of all the different n-cycles ($d_{\rm H} \simeq 1.201$).}
    \label{linearfit}
\end{figure*}

Due to the discreteness of the points in the stability diagram, the Hausdorff dimension is estimated via linear fits (see figure~\ref{linearfit}), exploiting the formula
\begin{equation}
    \delta_H = \lim_{r\rightarrow 0}\frac{\log{\left(N_b(r)\right)}}{\log{\left(r^{-1}\right)}},
\end{equation}
where $N_b(r)$ is the number of boxes having radius $r$ needed to cover the border of the regions.

In this scenario, limit-cycle solutions emerge. This kind of solution, after a transient time due to the dissipation, is trapped in stable and regular trajectories around $\varphi_\Delta = \pi$ or $0$. In our case, a limit-cycle solution is classified as an n-cycle, where the number n represents the number of the points of the trajectory having zero velocity, i.e. the number of times the trajectory intersects the zero velocity axis, on one side of the phase space. The n-cycles collected in the stability diagram are labelled with different colours (see figure~\ref{cycle}), while in figure~\ref{merged} the phase space portraits of six different n-cycles are shown. For instance, we analyze the trajectory in figure~\ref{figure:stable}d, having one intersection with zero velocity axis on the left and one on the right side of the point $\varphi_{\Delta} = \pi$: it represents a $1$-cycle. The system dynamics exhibits complex cyclic trajectories, such as 5-cycles (see figures~\ref{merged}b and~\ref{merged}e) and 9-cycles (see figures~\ref{merged}c and~\ref{merged}f). Furthermore, we find that n-cycles exhibit different topological properties and they can be distinguished through the turning number $t$ and the winding number $w$. The turning number represents the number of loops in a cyclic trajectory, which in our case is equal to the number of intersections n. The winding number counts the number of oscillations around a specific phase space point, which in our case can be $\left(0,0\right)$ or $\left(\pi,0\right)$. We note that n-cycle can occur both with $t = w$ and $t > w$ (see figure~\ref{merged}) for different values of the parameters.

We estimate numerically the number of intersections n, working on the phase space trajectories of the n-cycles. First, we estimate their period through the analysis of the component in the frequency space, i.e. through the peaks of the Discrete Fourier Transform of $\varphi_\Delta (t) - \bar{\varphi}_\Delta$, where $\bar{\varphi}_\Delta (t)$ is the mean value of the phase: the smallest positive frequency $\omega^*$ is inverted for the determination of the period $T^{*}=\frac{2\pi}{\omega^*}$: n is then determined by counting the times in which the velocity $\dot{\varphi}_\Delta (t)$ crosses the $\dot{\varphi}=0$ axis in one period $T^*$. The results of the classification are the n-cycle regions in figures~\ref{cycle}a-\ref{cycle}c, where each n-cycle is labelled with a different colour in the stability diagram. In particular, it is possible to identify three main subregions with $1$-cycle, $3$-cycle, and $5$-cycle. However, narrow stripes of limit-cycle solutions with higher numbers of nods occur for some specific values of the parameter $\bar{\epsilon}$, as reported in figures~\ref{cycle}b and~\ref{cycle}c. For instance, in a neighbourhood of $\bar{\epsilon} = 0.385$ a large band of $9$-cycles appears.

For the case $\dot{\varphi}_\Delta (0) \ne 0 $ we have obtained two basins of attraction (see figure~\ref{Fig5}), where a basin of attraction consists in the collection of all the initial conditions leading to a specific asymptotic attractor, obtained for constant values of the parameters.  The estimation of the Hausdorff dimensions of the basins of attraction follows the same procedure of the case $\dot{\varphi}_\Delta (0) = 0$ previously discussed.

\newcommand{\newblock}{}
\bibliographystyle{iopart-num.bst}
\bibliography{main.bib}

\end{document}